\documentclass[prb,aps,twocolumn,amsmath,amssymb,showpacs]{revtex4}
\usepackage{graphicx}
\usepackage{psfrag}
\usepackage{color}
\usepackage{subfigure}
\usepackage{amsmath}
\definecolor{dred}{rgb}{0.7,0.0,0.0}
\allowdisplaybreaks 
\usepackage{array}

\newcommand{\e}{{\rm e}}

\begin{document}

%
%

\title{Magnetic properties and Mott transition in the Hubbard model 
on the anisotropic triangular lattice}
\author{A. Yamada}
\affiliation{Department of Physics, Chiba University, Chiba 263-8522, Japan}

\date{\today}

\begin{abstract}

Magnetic properties and Mott transition are studied in the Hubbard model on the anisotropic triangular lattice described by 
two hopping parameters $t$ and $t'$ in different spatial directions using the variational cluster approximation. 
Taking into account N\'eel (AF), 120$^\circ$ N\'eel (spiral), and collinear (AFC) orderings, 
the magnetic phase diagram is analyzed at zero temperature and half-filling. 
We found six phases, AF-metal, AF-insulator, spiral, AFC, paramagnetic metal, and non-magnetic insulator, 
which is the candidate of spin liquid. 
Direct transitions from paramagnetic metal to AF insulator take place for $0.6 \lesssim t'/t \lesssim 0.8$, 
and non-magnetic insulator is realized between the paramagnetic metal and magnetic states for $0.8 \lesssim t'/t \lesssim 1.2 $. 
Around $ t'/t \simeq 1.2$, magnetic state (AFC or spiral) is realized above the paramagnetic metal, 
and as the on-site Coulomb repulsion $U$ increases, it changes to non-magnetic insulator. 
Implications for the $\kappa$-(BEDT-TTF)$_2$Cu$_2$(CN)$_3$ are discussed. 
As for the Mott transition, the structure of the self-energy in the spectral representation is studied in detail. 
As $U$ increases around the Mott transition point, single dispersion evolves in the spectral weights of the self-energy, 
which yields the Mott gap. 

\end{abstract}
 
 
\maketitle

%
%

\section{Introduction}

When kinetic and Coulomb repulsion energies are competing, low dimensional materials with geometric frustration 
exhibit rich phenomena like superconductivity with various pairing symmetries and purely paramagnetic insulator 
(spin liquid), which attract a lot of experimental and theoretical interests. 
The organic charge-transfer salts $\kappa$-(BEDT-TTF)$_2\mathrm{X}$~\cite{lefebvre00,shimizu03,kanoda3,manna} 
are good examples of such materials, where a transition from paramagnetic metal to spin liquid (Mott transition) 
has been detected with X=Cu$_2$(CN)$_3$.~\cite{kanoda3,manna} 
\begin{figure}
\includegraphics[width=0.47\textwidth,bb= 46.800300 203.098000 529.243000 424.367000]{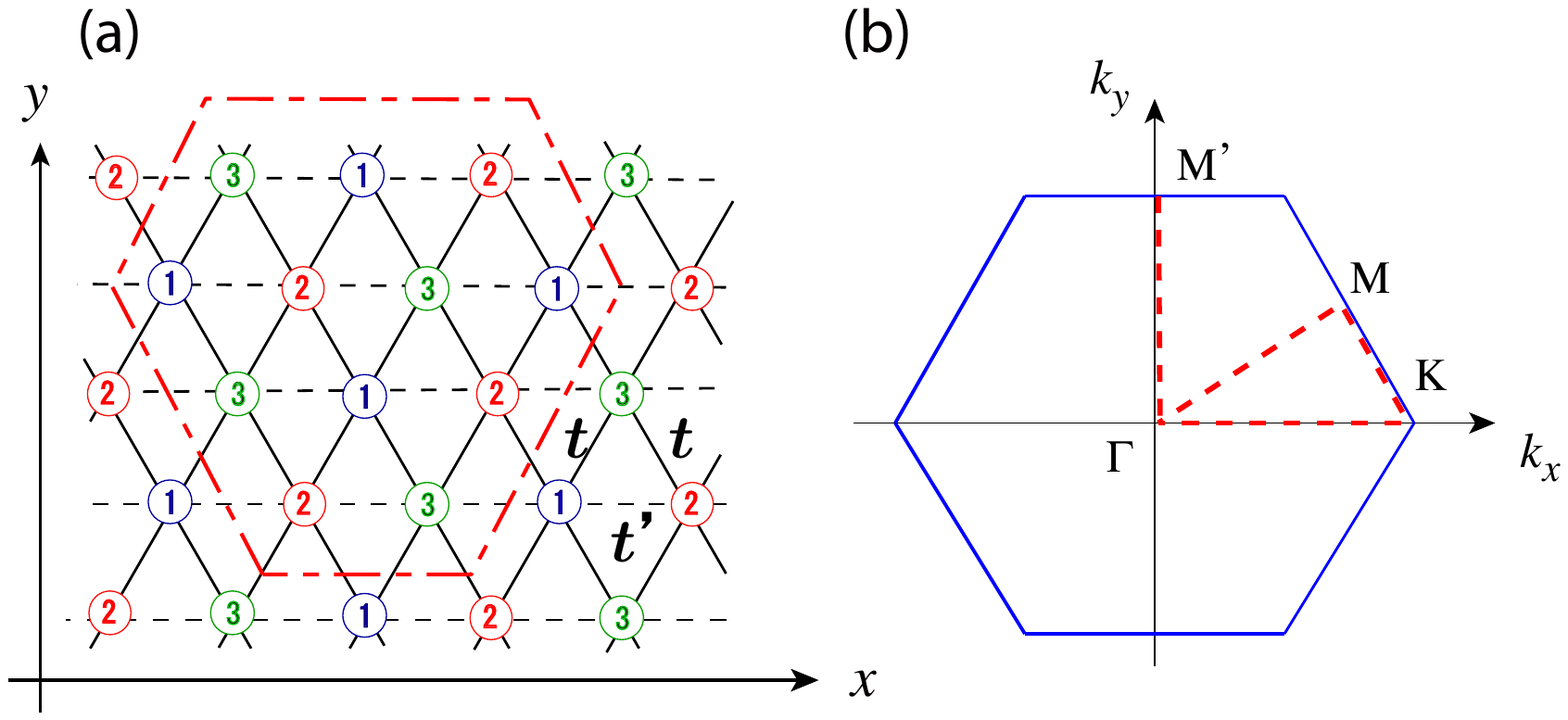}
\caption{(Color online) 
(a) Anisotropic triangular lattice. 
In our lattice geometry, the three sites $1,2$, and $3$ form an equilateral triangle of the unit length. 
The dash-dotted hexagon is the 12-site cluster used in our analysis. 
(b) The first Brillouin zone of the anisotropic triangular lattice. 
\label{fig:model}\\[-1.5em]}
\end{figure}

One of the important theoretical issues motivated by these experiments\cite{kanoda3,manna} is the possibility of spin liquid phase 
compatible with experiments in the Hubbard model on the anisotropic triangular lattice  
described by the hopping $t$, $t'$, and the on-site Coulomb repulsion $U$ (see Fig.~\ref{fig:model}(a)), 
which is a simple effective Hamiltonian of these materials and has been studied 
by various non-perturbative methods.
\cite{imada,kyung,senechal-af,clay,tocchio,Kokalj,kawakami,senechal,mila,antipov,watanabe,becca} 
\begin{figure}
\includegraphics[width=0.47\textwidth,bb = 121 126 486 239]{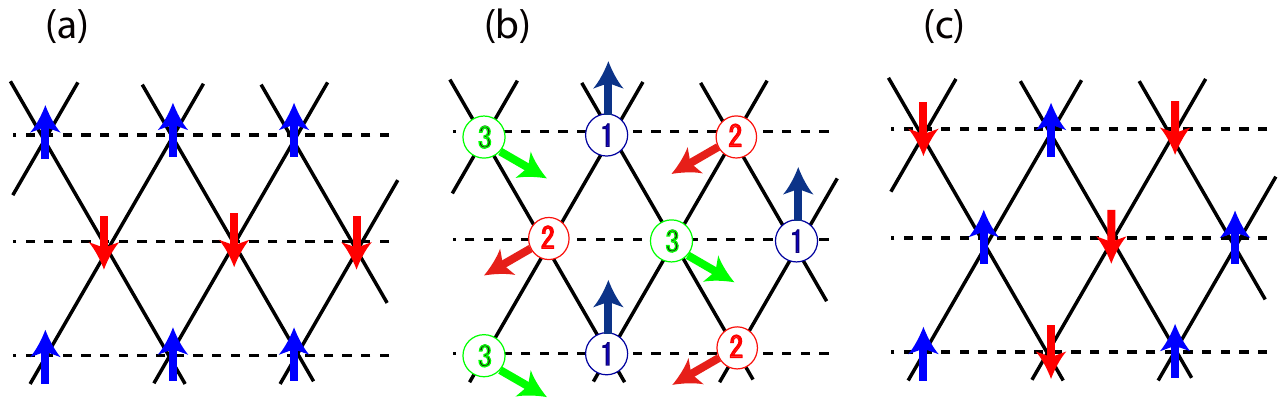}
\caption{(Color online)
Three magnetic orderings 
(a) N\'eel (AF), 
(b) 120$^\circ$ N\'eel (spiral), and 
(c) collinear (AFC) 
on the triangular lattice in Fig.~\ref{fig:model} (a).
\label{fig:spin-config}\\[-1.5em]}
\end{figure}

The earlier studies\cite{imada,kyung,senechal-af,clay,tocchio,Kokalj} reported that 
non-magnetic insulator, which is the candidate of spin liquid, is realized near the isotropic point $t'/t=1$, 
however 120$^\circ$N\'eel (see Fig.~\ref{fig:spin-config} (b)), 
which is the most relevant magnetic ordering around $t'/t=1$, was not considered in them. 
Taking into account 120$^\circ$ N\'eel, the isotropic case $t'/t=1$ is studied 
by the path integral renormalization group (PIRG),\cite{kawakami} 
variational cluster approximation (VCA),\cite{senechal} 
and using effective models,\cite{mila,antipov} where 
the PIRG study\cite{kawakami} predicted that non-magnetic insulator 
is realized for $7.4 \lesssim U/t \lesssim9.2$, 
and the other studies\cite{senechal,mila,antipov} 
also suggested the existence of non-magnetic insulator phase. 
For example, the VCA study\cite{senechal} argued that non-magnetic insulator is 
realized at least at $U/t =8$. 
More generic ranges of $t'/t$ are studied including 120$^\circ$ N\'eel by 
the variational Monte Carlo (VMC),\cite{watanabe,becca} 
where non-magnetic insulator is not found in the region $0 \leq t'/t \leq 1.2$ and $0 \leq U/t \leq 25$ 
in the older study\cite{watanabe}, while the later study\cite{becca} showed that non-magnetic insulator is not 
obtained at $t'/t =1$, but it is realized around $t'/t \simeq 0.85$ for $12 \lesssim U/t$. 
Therefore the conclusions about non-magnetic insulator in this model are 
very different not only depending on the approaches but also within the same approach. 

In this paper, we investigate the magnetic properties and Mott transition 
in the Hubbard model on the anisotropic triangular lattice 
using VCA,\cite{Senechal00,Potthoff:2003-1,Potthoff:2003} 
which is formulated based on a rigorous variational principle and exactly 
takes into account the short-range correlations. 
The 12-site cluster in the dash-dotted hexagon in Fig.~\ref{fig:model} (a) is used and 
the three magnetic orderings, N\'eel, 120$^\circ$ N\'eel, and collinear in Fig.~\ref{fig:spin-config}, 
which are referred to as AF, spiral, and AFC hereafter, are considered 
to analyze the magnetic phase diagram at zero temperature and half-filling. 

We found six phases, AF-metal, AF-insulator, spiral, AFC, paramagnetic metal, and non-magnetic insulator, 
which is the candidate of spin liquid. 
AF-metal is realized for $ t'/t \lesssim 0.6$ and relatively small $U$. 
Direct transitions from paramagnetic metal to AF-insulator take place for $0.6 \lesssim t'/t \lesssim 0.8$ around 
$U = 5 \sim 6$. 
For $0.8 \lesssim t'/t \lesssim 1.2$ paramagnetic metal changes to non-magnetic insulator 
at $U/t \simeq 6$, thus the (purely paramagnetic) Mott transition takes place there, and this non-magnetic insulator become 
magnetic state at $U/t \simeq 8$. 
Around $ t'/t \simeq 1.2$, magnetic state (AFC or spiral) is realized above the paramagnetic metal, 
and it changes to non-magnetic insulator as $U$ increases.

Implications of our analysis for experiments on the organic charge-transfer salts $\kappa$-(BEDT-TTF)$_2\mathrm{X}$ 
are discussed. 
As for the Mott transition, the structure of the self-energy in the spectral representation is studied in detail. 
As $U$ increases around the Mott transition point, single dispersion evolves in the spectral weights of the self-energy, 
which gives rise to the splitting of the non-interacting band into the upper and lower Hubbard bands. 

%
%
\section{Hubbard model on the anisotropic triangular lattice}

The Hamiltonian of the Hubbard model on the anisotropic triangular lattice reads 
\begin{align}
H =& -\sum_{i,j,\sigma} t_{ij}c_{i\sigma }^\dag c_{j\sigma}
+ U \sum_{i} n_{i\uparrow} n_{i\downarrow} - \mu \sum_{i,\sigma} n_{i\sigma},
\label{eqn:hm}
\end{align}
where $t_{ij}=t$ for the solid lines and $t_{ij}=t'$ for the dashed lines in Fig.~\ref{fig:model} (a), $U$ is 
the on-site Coulomb repulsion, and $\mu$ is the chemical potential. 
The annihilation (creation) operator for an electron at site $i$ with spin $\sigma$ is denoted as 
$c_{j\sigma}$ ($c_{i\sigma }^\dag$) and $n_{i\sigma}=c_{i\sigma}^\dag c_{i\sigma}$. 
The energy unit is set as $t=1$ hereafter.

%
%

In our analysis we use VCA,\cite{Senechal00,Potthoff:2003-1,Potthoff:2003} which 
is an extension of the cluster perturbation theory\cite{Senechal00} based on the 
self-energy-functional approach.\cite{Potthoff:2003} 
This approach uses the rigorous variational principle 
$\delta \Omega _{\mathbf{t}}[\Sigma ]/\delta \Sigma =0$ for the thermodynamic grand-potential 
$\Omega _{\mathbf{t}}$ written in the form of a functional of the self-energy $\Sigma $ as
\begin{equation}
\Omega _{\mathbf{t}}[\Sigma ]=F[\Sigma ]+\mathrm{Tr}\ln(-(G_0^{-1}-\Sigma )^{-1}).
\label{eqn:omega}
\end{equation}%
In Eq. (\ref{eqn:omega}), $G_0$ is the non-interacting Green's function of $H$, $F[\Sigma ]$ is the Legendre 
transform of the Luttinger-Ward functional,\cite{lw} and the index $\mathbf{t}$ denotes the explicit dependence of 
$\Omega _{\mathbf{t}}$ on all the one-body operators in $H$. 
The variational principle $\delta \Omega _{\mathbf{t}}[\Sigma ]/\delta \Sigma =0$ leads to the Dyson's equation. 
Eq. (\ref{eqn:omega}) gives the exact grand potential for the exact self-energy of $H$, 
which satisfies Dyson's equation. 

All Hamiltonians with the same interaction part share the same functional form of $F[\Sigma ]$, and using that property 
$F[\Sigma ]$ can be evaluated for the self-energy of a simpler Hamiltonian $H'$ by exactly solving it, though 
the space of the self-energies where $F[\Sigma ]$ is evaluated is now restricted to that of $H'$. 
In VCA, one uses for $H'$ a Hamiltonian formed of clusters that are disconnected by removing hopping terms 
between identical clusters that tile the infinite lattice. 
For $H'$, the grand potential is expressed as a functional of $\Sigma$ as
\begin{equation}
\Omega' _{\mathbf{t'}}[\Sigma ]=F[\Sigma ]+\mathrm{Tr}\ln(-(G'_0{}^{-1}-\Sigma )^{-1}),
\label{eqn:omegaprime}
\end{equation}%
where $G'_0$ is the non-interacting Green's function of $H'$ and $\mathbf{t}'$ denotes all the one-body operators in $H'$. 
In Eqs. (\ref{eqn:omega}) and (\ref{eqn:omegaprime}), $F[\Sigma ]$ is the same for a given $\Sigma$ since the interaction 
part is the same for $H$ and $H'$, therefore subtracting Eq. (\ref{eqn:omegaprime}) from Eq. (\ref{eqn:omega}), 
we obtain 
\begin{eqnarray}
\Omega _{\mathbf{t}}[\Sigma ]= \Omega' _{\mathbf{t'}}[\Sigma ] &+& \mathrm{Tr}\ln(-(G_0^{-1}-\Sigma )^{-1})
\nonumber \\
&-&\mathrm{Tr}\ln(-(G'_0{}^{-1}-\Sigma )^{-1}),
\label{eqn:omega3}
\end{eqnarray}%
which is a functional relation between $\Omega _{\mathbf{t}}[\Sigma ]$ and $\Omega' _{\mathbf{t'}}[\Sigma ]$. 
In Eq. (\ref{eqn:omega3}) $\Omega' _{\mathbf{t'}}[\Sigma ]$ and $\Sigma$ are exactly computed for $H'$ by exactly solving it, 
thus $\Omega _{\mathbf{t}}[\Sigma ]$ is evaluated for the exact self-energy of $H'$, and becomes 
a function of $\mathbf{t}'$ expressed as 
\begin{equation}
\Omega _{\mathbf{t}}(\mathbf{t}')=\Omega' _{\mathbf{t'}} - \int_C{\frac{%
d\omega }{2\pi }} \e^{ \delta \omega} \sum_{\mathbf{K}}\ln \det \left(
1+(G_0^{-1}\kern-0.2em -G_0'{}^{-1})G'\right),
\nonumber
\end{equation}%
where $\Omega' _{\mathbf{t'}}$ is the exact grand potential of $H'$ and 
the functional trace has become an integral over the diagonal variables 
(frequency and super-lattice wave vectors) of the logarithm of the determinant over intra-cluster indices. 
The frequency integral is carried along the imaginary axis and $\delta \rightarrow + 0$. 
\begin{figure}
\includegraphics[width=0.48\textwidth,bb = 128 300 467 541]{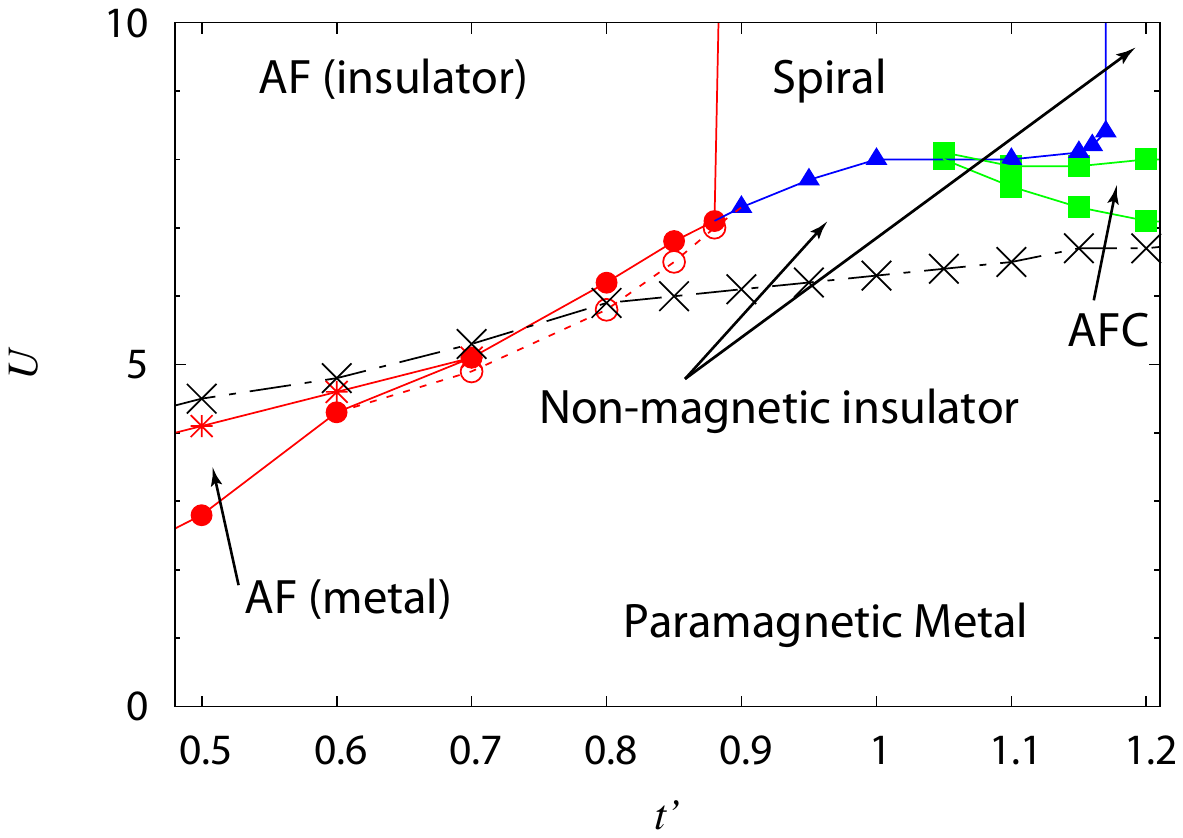}\\[-0.5em]
\caption{
(Color online)
Phase diagram of the Hubbard model on the anisotropic triangular lattice
at zero temperature and half-filling as a function of $t'$ and $U$ obtained by VCA on 12D cluster. 
Lines are guides to the eye.
The filled circles, triangles, and squares correspond to AF, spiral, and AFC transition points, and 
the two asterisks at $t'=0.5$ and $0.6$ denote the transition points from AF insulator to AF metal. 
Energetically disfavored magnetic solutions are obtained inside the non-magnetic phase between the unfilled and filled circles. 
The crosses are the Mott transition points computed assuming that no magnetic order is allowed. 
The line between AF and spiral phases will actually not be a phase boundary 
and AF will gradually change to spiral through complicated (probably incommensurate) orderings. 
Energetically disfavored magnetic solutions between the unfilled and filled circles may be ground states with these 
complicated magnetic orderings. 
\label{fig:phase-diagram}
}
\end{figure}

The variational principle $\delta \Omega _{\mathbf{t}}[\Sigma ]/\delta \Sigma =0$ is reduced to the stationary condition 
$\delta \Omega _{\mathbf{t}}(\mathbf{t}') /\delta \mathbf{t}' = 0$, and its solution and the 
exact self-energy of $H'$ at the stationary point, denoted as $\Sigma^{*}$, are the approximate grand-potential 
and self-energy of $H$ in VCA. Physical quantities, such as expectation values of one-body operators, 
are evaluated using the Green's function $G_0{}^{-1}-\Sigma^{*} $. 
In VCA, the restriction of the space of the self-energies $\Sigma$ into that of $H'$ 
is the only approximation involved and short-range correlations within the cluster are exactly taken into account 
since $H'$ is solved exactly. 
A possible symmetry breaking is investigated by including in $H'$ the corresponding Weiss field that will be 
determined by minimizing the grand-potential $\Omega_{\mathbf{t}}$. 
\begin{figure}
\includegraphics[width=0.44\textwidth,bb = 187 66 514 559]{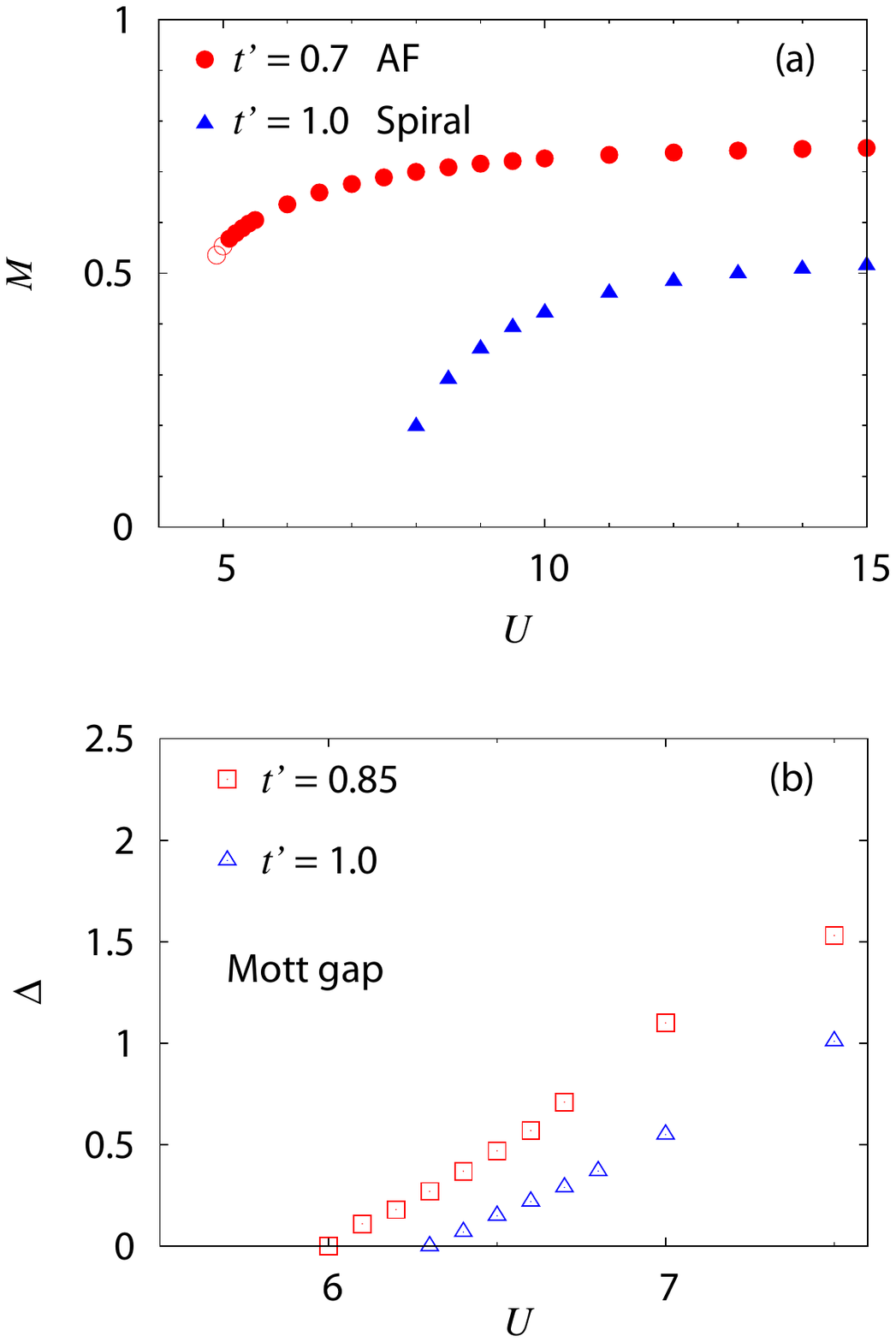}\\
\caption{
(Color online)
(a) The magnetic order parameters $M$ for AF (filled-circles) at $t'=0.7$ and spiral (filled-triangles) at $t'=1.0$ 
as functions of $U$. The unfilled marks correspond to the energetically disfavored solutions.
(b) The Mott gaps at $t' = 0.85$ (squares) and $t' = 1.0$ (triangles) computed as functions of $U$ 
assuming that no magnetic order is allowed. 
\label{fig:order-tp}\\[-2.2em]
}
\end{figure}

In our analysis, the 12-site cluster of the diamond shape in the dash-dotted hexagon in Fig.~\ref{fig:model}(a), 
which is referred to as 12D hereafter, is used to set up the cluster Hamiltonian $H'$. 
This cluster treats the three sub-sites 1,2, and 3 on the same footing, and has even number of sites 
so that energy eigenstates with total $S_z=0$, which is satisfied for the non-magnetic ground states of the infinite system $H$, 
are included as candidates of the grand states of $H'$. 
To study the magnetic orderings AF, spiral, and AFC, the Weiss field 
\begin{eqnarray}
H_{\rm AF}&=& h_{\rm M}\sum_{i} {\bf e}_{a_i} \cdot {\bf S}_i
\label{eqn:weiss}
\end{eqnarray}
with the spin operator ${\bf S}_i = c_{i\alpha}^\dag \sigma_{\alpha\beta}c_{i\beta}$ is included into $H'$, 
where the index $a$ specifies the site in the unit cell in the sub-lattice 
formalism, and $a=1,2,3$ for spiral, and $a=1,2$ for AF and AFC.
The unit vectors ${\bf e}_{1,2,3}$ are oriented at 120$^\circ$ of each other for spiral, 
and ${\bf e}_{1}= -{\bf e}_{2} $ for AF and AFC according to these spin orderings (see Fig.~\ref{fig:spin-config}). 
In our analysis the pitch angle of spiral order is fixed to be 120$^\circ$ even for $t' \neq t$. 

In the stationary point search of $\Omega(\mu', h_{\rm M})$, which we denote as the grand-potential per site, 
the Weiss field parameter $h_{\rm M}$ and the cluster chemical potential $\mu'$ in $H'$ 
are treated as the variational parameters, where 
$\mu'$ should be included for the thermodynamic consistency.\cite{aichhorn} 
During the search, the chemical potential of the system $\mu$ is also adjusted so that the electron 
density $n$ is equal to 1 within 0.1\%. 
In general, a stationary solution with $h_{\rm M} \neq 0$ corresponding to the magnetically ordered state and 
that with $h_{\rm M} = 0$ corresponding to the paramagnetic state (PM) 
are obtained, and the energies per site $E=\Omega+\mu n$ are compared 
for AF, spiral, AFC, and PM to determine the ground state. 
The density of state per site 
\begin{eqnarray}
D(\omega)= \lim_{\eta \rightarrow 0}  \int
{\frac{%
d^2 k }{(2\pi)^2 }}\frac{1}{n_c}\sum_{\sigma, a=1}^{n_c}\{ -\frac{1}{\pi} \mathrm{Im}G_{a\sigma}(k, \omega+i\eta) \}
\label{eqn:dos}
\end{eqnarray}
is also calculated to examine the gap, where 
$n_c$ is the number of the sites in the unit cell in the sense of the sub-lattice formalism 
($n_c = 3$ for spiral, $n_c = 2$ for AF and AFC, and $n_c = 1$ for PM), 
and the $k$ integration is over the corresponding Brillouin zone. 
In Eq. (\ref{eqn:dos}), $\eta \rightarrow 0$ limit is evaluated using the standard extrapolation method 
by calculating $D(\omega)$ for $\eta =0.1$, $0.05$, and $0.025$. 
The numerical error after this extrapolation is estimated to be of order $10^{-3}$, 
so the gap is identified as the region of $\omega$ around $\omega \simeq 0 $ 
where the extrapolated $D(\omega)$ is less than $10^{-2}$. 
We also compute the magnetic order parameter per site 
\begin{eqnarray}
M&=& \frac{1}{n_c}\sum_{a=1}^{n_c}  {\bf e}_a \cdot \langle {\bf S}_a  \rangle\nonumber
\end{eqnarray}
and the double occupancy per site 
\begin{eqnarray}
D_{\rm occ}= \frac{1}{n_c}\sum_{a=1}^{n_c} \langle n_{a\uparrow } n_{a\downarrow } \rangle = \frac{dE}{dU}, \nonumber 
\end{eqnarray}
where $ \langle {\bf S}_a \rangle $ and $\langle n_{a\uparrow } n_{a\downarrow } \rangle$ 
are the expectation values of ${\bf S}_a$ and 
$n_{a\uparrow } n_{a\downarrow }$, respectively. 
\begin{figure}
\includegraphics[width=0.47\textwidth, bb = 144 300 465 541]{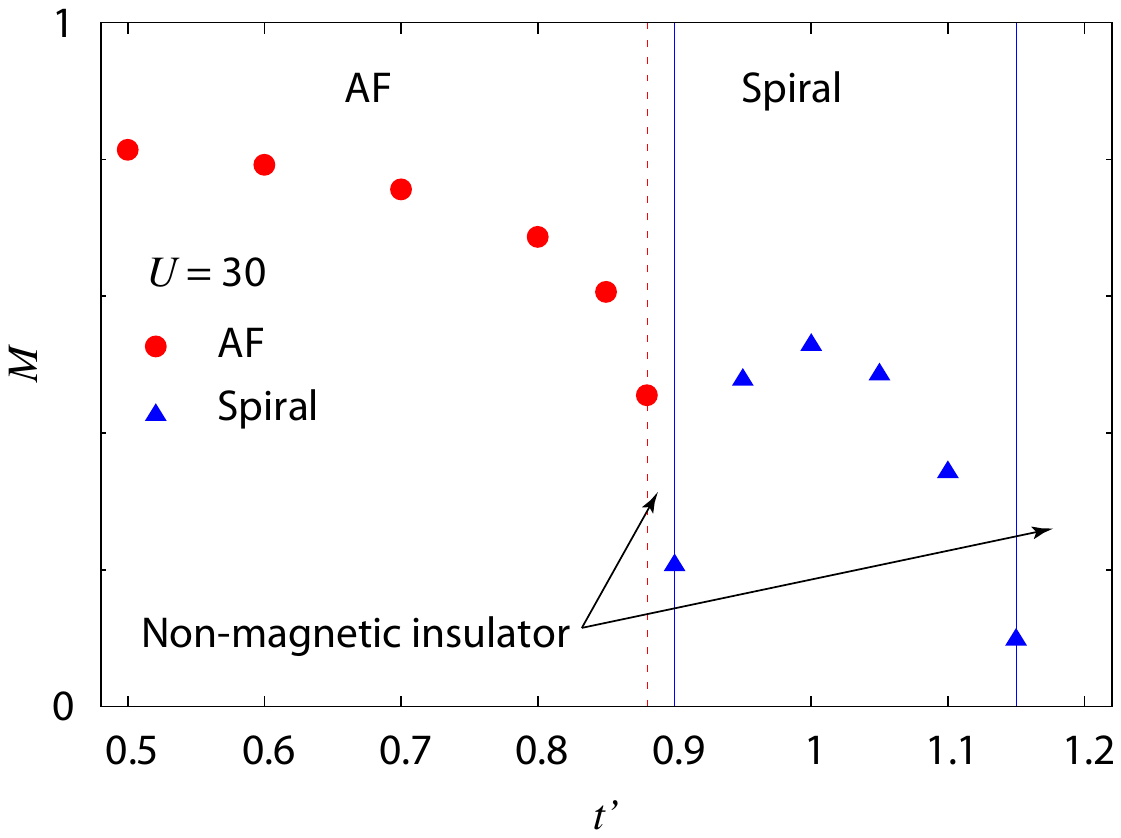}\\
\caption{
(Color online)
The magnetic order parameters $M$ for AF (filled-circles) and spiral (filled-triangles) at $U=30$ as functions of $t'$. 
Magnetic solutions including energetically disfavored ones are not found 
in the region around $t' = 0.89$ and $1.15 < t'$. 
\label{fig:order-u30-15}\\[-2.2em]
}
\end{figure}

\section{Phase diagram}

\begin{figure}
\includegraphics[width=0.47\textwidth, bb= 34 41 539 387]{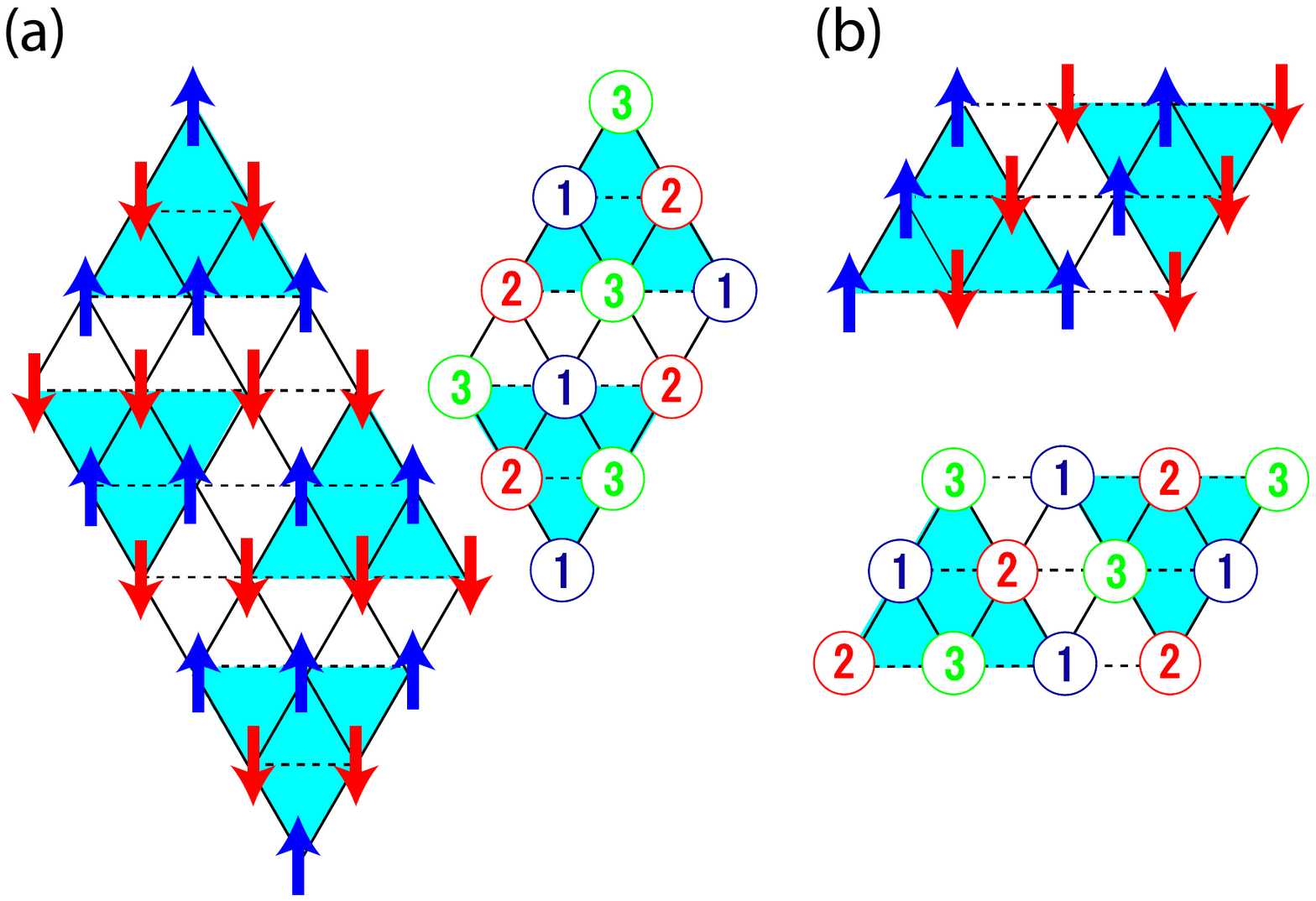}\\
\caption{
(Color online)
The 6-site clusters (shaded triangles) used in VCA.  
Two or four 6-site clusters are combined as in (a) $ t' \leq 1$ and (b) $ 1 \leq t'$ 
to recover the lattice geometry in the presence of the magnetic orderings. 
\label{fig:6sites}
}
\end{figure}
Fig.~\ref{fig:phase-diagram} shows the phase diagram at zero temperature and half-filling 
obtained by VCA on 12D cluster, where 
the filled circles, triangles, and squares correspond to the AF, spiral, and AFC transition points and 
the two asterisks at $t'=0.5$ and $0.6$ denote the transition points from AF insulator to AF metal. 
Energetically disfavored AF solutions are obtained between the unfilled and filled circles, 
and except this region, energetically disfavored magnetic solutions (AF, spiral, or AFC) are not obtained 
inside the non-magnetic phases. 
The crosses are the Mott transition points computed assuming that no magnetic order is allowed (i.e. $h_{\rm M}$ = 0). 
The magnetic order parameters and Mott gaps are shown in Fig.~\ref{fig:order-tp}. 
Near $t'\simeq 0.5$, the magnetic transition point is relatively low and the similar phase diagram is obtained also in the 
mean field study.\cite{kino}

Non-magnetic insulator is realized above the Mott transition line for 
$0.8 \lesssim t' \lesssim 1.2$, and it changes into magnetic states as $U$ increases. 
Non-magnetic insulator phase exists also for $1.15 \lesssim t'$ and $ 8 \lesssim U$ above AFC. 
These non-magnetic insulator phases are the candidates of spin liquid. 
In Fig.~\ref{fig:phase-diagram} the line between AF and spiral phases will actually not be a phase boundary 
and AF will gradually change to spiral via complicated (probably incommensurate) magnetic orderings 
as $t'$ increases. These more complicated states will give local minimums (i.e. stationary solutions) 
with our choice of the Weiss fields given by Eq. (\ref{eqn:weiss}) since they will have orderings 
similar to AF or spiral, but their actual energies might be lower than the values 
computed with our Weiss fields. 
So these states may appear, not only as ground states, but also as energetically disfavored solutions, 
which may turn out to be ground states by more appropriate choice of Weiss fields. 
However, these energetically disfavored solutions are obtained only in the restricted 
region between the filled and unfilled circles in Fig.~\ref{fig:phase-diagram}, and most of the non-magnetic insulator phase 
in that phase diagram remains stable even if these energetically disfavored solutions become 
ground states with complicated magnetic orderings. 

For $1.05 \lesssim t' \lesssim 1.15$ and $U \simeq 8$, AFC appears in addition to spiral so the magnetic orderings in this 
region will have both spiral and AFC like features. 
Around $t' \simeq 1.2$ spiral disappears and AFC remains, which suggests 
that in the region $1.2 \leq t'$ the correlations along $t'$ direction become important, probablly 
because the system begins to evolve towards weakly coupled one-dimensional chains, 
so this region is not investigated in this study. 

In Fig.~\ref{fig:phase-diagram}, the transition from the non-magnetic to magnetic states 
is of the second order except for $0.6 < t' < 0.9$ since there exists no energetically disfavored magnetic solution 
outside the magnetic phase except this range. 
If the energetically disfavored solutions in this range are magnetic ground states as is discussed above, 
the magnetic transition is of the second order for all $t'$. 
The Mott transition is of the second order since there is no energetically disfavored 
paramagnetic solution near the transition line and the Mott gap closes continuously at the transition point as is shown 
in Fig.~\ref{fig:order-tp} (b). 
\begin{figure}
\includegraphics[width=0.47\textwidth,trim = 0 0 0 0,bb= 159 119 546 313]{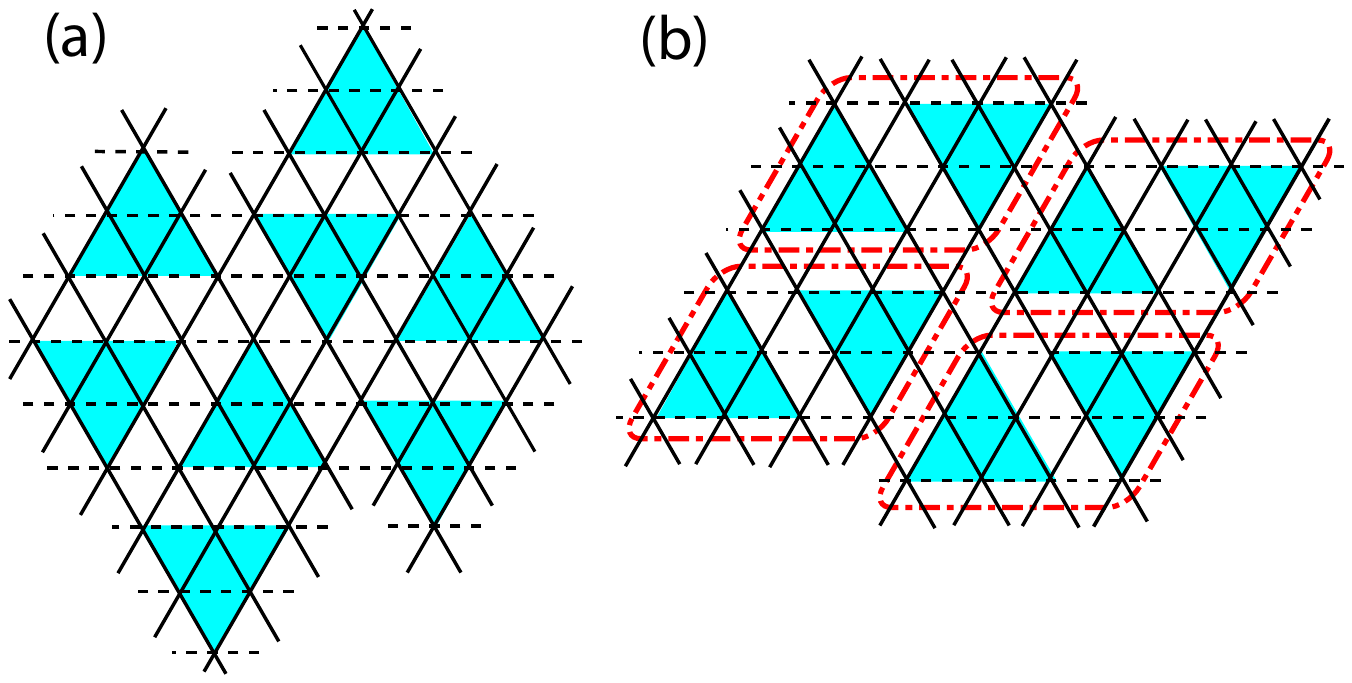}\\
\caption{
(Color online)
The tilings of the triangular 6-site clusters used for (a) $ t' \leq 1$ and (b) $ 1 \leq t'$. 
The tilings (a) and (b) are equivalent at $t'/t=1$. 
The dash-dotted parallelograms in (b) are the 3$\times$4 clusters and their tiling 
used to study the region $ t' = 1.2$ in VCA. 
\label{fig:3x4tiling}\\[-2.2em]
}
\end{figure}

We have investigated the magnetic properties up to $U =30$. 
Fig.~\ref{fig:order-u30-15} shows the AF (filled-circles) and spiral (filled-triangles) 
order parameters $M$ at $U=30$ as functions of $t'$. 
The non-magnetic insulator phase above AFC in Fig.~\ref{fig:phase-diagram} persists up to large $U$. 
The magnetic phase (AF and spiral) covers whole the region $ t' \leq 1.15$ up to $U \lesssim 15 \sim 20$. 
For $20 \lesssim U$ there appears a very narrow slit of non-magnetic insulator phase 
around $t' = 0.89$ (between the dotted and full lines) 
where no magnetic solution (including energetically disfavored one) is obtained. 

Our Fig.~\ref{fig:order-u30-15} is qualitatively very similar to Fig.~4 in Ref.~\onlinecite{trumper} 
obtained by the spin wave theory taking into account, not only AF, spiral, and AFC but also 
general magnetic ordering vectors including incommensurate ones. 
In that figure $\eta $ is equal to $1/(1 +t'^2/t^2 )$ in the large $U$ limit and 
their collinear corresponds to the ordering similar to our AF. 
The point where the magnetic order parameter vanishes is obtained in the region $t' < 1$ 
also in the spin wave theory considering general magnetic ordering vectors. 
A narrow non-magnetic phase is found near $ t'/t \simeq 0.9$ in the modified spin wave theory 
complemented by exact diagonalizations.\cite{hauke}
\begin{figure}
\includegraphics[width=0.47\textwidth, bb = 144 306 465 541]{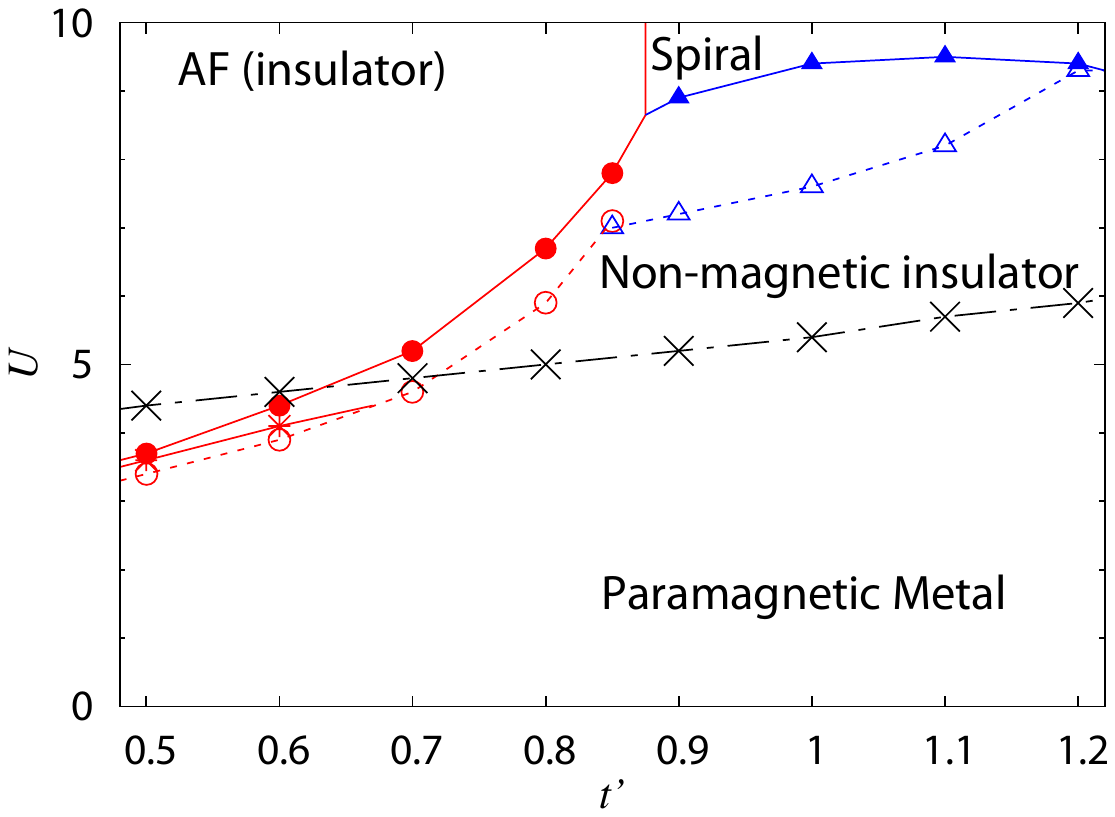}\\[-0.5em]
\caption{
(Color online)
Phase diagram of the Hubbard model on the anisotropic triangular lattice
at zero temperature and half-filling as a function of $t'$ and $U$ obtained by VCA on the triangular 6-site cluster. 
Lines are guides to the eye.
The filled circles and triangles correspond to the AF and spiral transition points and 
crosses are the Mott transition points obtained assuming that no magnetic order is allowed. 
The two asterisks at $t'=0.5$ and $0.6$ denote the points where AF gap closes for energetically disfavored solutions. 
Energetically disfavored magnetic (AF and spiral) solutions are obtained between the unfilled and filled marks. 
\label{fig:phase-diagram-6}
}
\end{figure}

Here we remark that only the three magnetic orderings AF, spiral (of 120$^\circ$ pitch angle), and AFC 
(i.e., only the three magnetic ordering vectors in the sense of the spin wave) are considered in our analysis 
and our Weiss fields will not be able to detect magnetic orderings very different from the above three. 
For example, magnetic orderings whose modulation period does not fit into the cluster size 
are not investigated well in VCA and we can not exclude the possibility that such magnetic orderings are realized in the 
non-magnetic phase found in Figs.~\ref{fig:phase-diagram} and~\ref{fig:order-u30-15}. 
Later we further compare quantitatively our results in Fig.~\ref{fig:order-u30-15} with recent 
non-perturbative analyses of the Heisenberg model 
which studied the possibility of general spiral order in addition to our 120$^\circ$ spiral. 

Next we consider the cluster size dependence of our phase diagram. 
In general, ordered phase shrinks as the cluster size 
increases since spatial fluctuations, which destabilize ordered states, are better simulated on larger clusters. 
However, when magnetic ordering is rather suppressed, wave functions with small magnetic fluctuations 
realized on larger clusters play an important role to examine near the true minimum of the effective potential, 
thus ordered phase may become wider as the cluster size increases. 
The critical interaction strength of the metal-insulator transition $U_{\rm MI}$ 
(denoted by the crosses in Fig.~\ref{fig:phase-diagram}) decreases as the cluster size increases, 
since the average kinetic energies are lower on larger cluster and metal is stabilized. 

To see in detail the cluster size dependence, first we analyze the phase diagram 
by VCA on the triangular 6-site cluster. 
To study AF, spiral, and AFC, we combined two and four 6-site clusters as depicted in Fig.~\ref{fig:6sites} 
to recover the lattice geometry in the presence of the magnetic orderings and then tiled the infinite lattice 
with them as in Fig.~\ref{fig:3x4tiling}. 
In these cases the Green's function $G'$ of the combined cluster is given by 
\begin{eqnarray}
G'^{-1}&=& \sum_{i} {G'}_i^{-1} + \tilde{t}
\end{eqnarray}
where ${G'}_i$ is the exact Green's function on each 6-site cluster (the site and spin indices suppressed) 
and $\tilde{t}$ is the hopping matrix linking these 6-site clusters. 
The Hamiltonian on the triangular 6-site cluster is exactly diagonalized in all cases therefore 
the correlations within the 6-site clusters (shaded triangle clusters in Figs.~\ref{fig:6sites} and ~\ref{fig:3x4tiling}) 
are exactly taken into account. 

Fig.~\ref{fig:phase-diagram-6} shows the phase diagram at zero temperature and half-filling 
obtained by VCA on the 6-cluster, where the filled circles and triangles correspond to the AF and spiral transition points. 
AFC phase is not realized in this parameter space. 
Energetically disfavored AF and spiral solutions are obtained between the filled and unfilled marks. 
The two asterisks at $t'=0.5$ and $0.6$ denote the points where AF gap closes for energetically disfavored solutions. 
The crosses are the Mott transition points computed assuming that no magnetic order is allowed. 

Comparing Fig.~\ref{fig:phase-diagram} and Fig.~\ref{fig:phase-diagram-6}, 
the magnetic transition point $U_{\rm c}$ is larger on the 6-site cluster for $t' \lesssim 1.1$, contrary to the general 
arguments on the cluster size dependence, and energetically disfavored solutions are obtained in rather 
wide region of the parameter space, while they are not obtained except the very restricted region on 12D cluster. 
Therefore the true minimum of the effective potential would not be yet simulated well on the 6-site cluster and 
the magnetic solutions turn out to be energetically disfavored or even can not be found in Fig.~\ref{fig:phase-diagram-6}. 
Mott transition line shifts upwards as the cluster size increases, which is consistent with the general argument 
on the cluster size dependence. 
Taking into account the energetically disfavored solutions, almost all the phases found on 12D cluster are also 
observed on the 6-site cluster and the shape of the non-magnetic 
insulator phase between the dotted and dash-dotted lines 
in Fig.~\ref{fig:phase-diagram-6} is more or less the same to 12D cluster for $0.7 \lesssim t' \lesssim 1.1$ except the 
upward shifts.
\begin{table}
\caption{
The Mott transition points and regions of magnetic phase at $t' = 1.2$ obtained by VCA on 3$\times$4 cluster 
in Fig.~\ref{fig:3x4tiling} (b) and 12D cluster. 
}
\vskip0.2cm
\begin{tabular}{ccc} \hline\hline
                         & \,\,\,$U_{\rm MI}$ \,\,\, & \,\,\, Magnetic phase           \,\,\, \\
\,\,\, 3$\times$4 \,\,\, & \,\,\, 6.0 \,\,\,         & \,\,\, $ 5.4 \leq U \leq 7.0 $  \,\,\, \\
\,\,\, 12D        \,\,\, & \,\,\, 6.7 \,\,\,         & \,\,\, $ 7.1 \leq U \leq 8.0 $  \,\,\, \\ \hline\hline
\end{tabular} 
\label{table:mott-3x4}
\end{table}

Next we consider the region $t' \simeq 1.2$. 
In this region, spiral phase disappears while AFC phase remains on 12D cluster, which suggests that 
the correlations along $t'$ direction become important probably because the system begins to evolve 
towards weakly coupled one-dimensional chains. 
Therefore we analyze the magnetic phase diagram also using the 3$\times$4 cluster in the dash-dotted parallelogram 
in Fig.~\ref{fig:3x4tiling} (b), which contains three 4-site chains in $t'$ direction. 
The infinite lattice is tiled as depicted there, and only spiral (of 120$^\circ$ pitch angle) and AFC are investigated. 
The correlations within the 3$\times$4 cluster are exactly taken into account in this study. 
The region of magnetic phase and Mott transition point $U_{\rm MI}$ are computed at $t' = 1.2$ in Table~(\ref{table:mott-3x4}), 
comparing with the results of 12D cluster. 
On 3$\times$4 cluster, AFC solutions are also obtained in addition to spiral in some region of the magentic phase and 
spiral has lower energy, while only AFC solutions are obtained on 12D. 
These subtle differences may arise due to the symmetry of the clusters used because 
the magnetic ordering in this region will have both spiral and AFC features. 
Even though there remains some cluster shape dependence, the general features are the same for these two clusters. 
Magnetic phase is realized between the paramagnetic metal and non-magnetic insulator at $t' = 1.2$, and 
non-magentic insulator above the magnetic phase persists up to large $U$. 
As was remarked previously, we can not exclude the possibility that magnetic orderings not approximated well by 
AFC or spiral (of 120$^\circ$ pitch angle), 
e.g. incommensurate spiral order suggested in the spin wave theory,\cite{trumper} are realized at $ t' = 1.2$.

Next we examine the cluster size dependence 
by comparing our results with the previous VAC study\cite{senechal} on the isotropic triangular lattice ($t'/t =1$). 
The general considerations on the cluster size dependence suggest 
that the magnetic phase shrinks and Mott transition line will shift upwards in the 
thermodynamic limit in Fig.~\ref{fig:phase-diagram}. 
As for the Mott transition, $U_{\rm MI} \simeq 6.7 $ in the thermodynamic limit,\cite{senechal} 
which is in fact slightly larger than our value $U_{\rm MI} = 6.3$ on 12D cluster, 
and gives us an estimate of the upper limit of the upward shift 
of our Mott transition line in Fig.~\ref{fig:phase-diagram}. 
As for spiral order, it is argued in Ref.~\onlinecite{senechal} that 
spiral order is absent in the thermodynamic limit at least at $U=8$, 
which is equal to our critical interaction strength $U_{\rm c} = 8.0$. 
This implies that our magnetic transition line in Fig.~\ref{fig:phase-diagram} can not shift downwards and 
should shift upwards. 
Therefore in the thermodynamic limit, the Mott transition line does not shift upwards by more than about 0.4 
and the magnetic transition line shifts upwards in Fig.~\ref{fig:phase-diagram}, 
thus the main features of the phase diagram remain almost the same except the small upwards shift, 
in particular non-magnetic insulator is realized above the paramagnetic metal for $0.8 \lesssim t'/t \lesssim 1.2$.
(As was shown in the above discussions, our VCA study is consistent with the previous study.\cite{senechal})

Next we compare our results in Fig.~\ref{fig:order-u30-15} with 
the analyses of the Heisenberg model on the triangular lattice.\cite{harada,heisenberg-order,hauke,heisenberg2,heisenberg3} 
Our value of the order parameter, $M = 0.528$ at $t'/t=1$ and $U=30$, 
lies almost at the center of the range\cite{harada} $ 0.464 \leq M \leq 0.596 $ (computed from $M_c$), 
and is larger than $M \simeq 0.38 \sim 0.40$ in Refs.~\onlinecite{heisenberg-order}, which suggests 
that our VCA analysis on 12D cluster still exaggerates the tendency of system to order 
and our magnetic transition line in Fig.~\ref{fig:phase-diagram} shifts upwards in the thermodynamic limit. 

In the Heisenberg model,  
it is argued in Ref.~\onlinecite{harada} that the incommensurate spiral order persists at least up to 
$t'/t = 1.2$, which is slightly larger than our value $t'/t = 1.15$ obtained taking into account only the 120$^\circ$ spiral. 
Non-magnetic insulator is predicted\cite{hauke,heisenberg2} around $ t'/t \sim 1.2$ though the ranges of 
$t'/t$ vary depending on the analyses. 
Our results in Fig.~\ref{fig:order-u30-15} are consistent with 
these analyses.\cite{hauke,harada,heisenberg-order,heisenberg2}
Contrary to these results,\cite{hauke,harada,heisenberg-order,heisenberg2} 
non-magnetic insulator is not found for $1 \le t'/t $ in Refs.~\onlinecite{heisenberg3}. 
So conclusions on non-magnetic insulator state in the Heisenberg model seem to be still controversial. 
\begin{figure}
\includegraphics[width=0.47\textwidth,bb = 130 305 463 537]{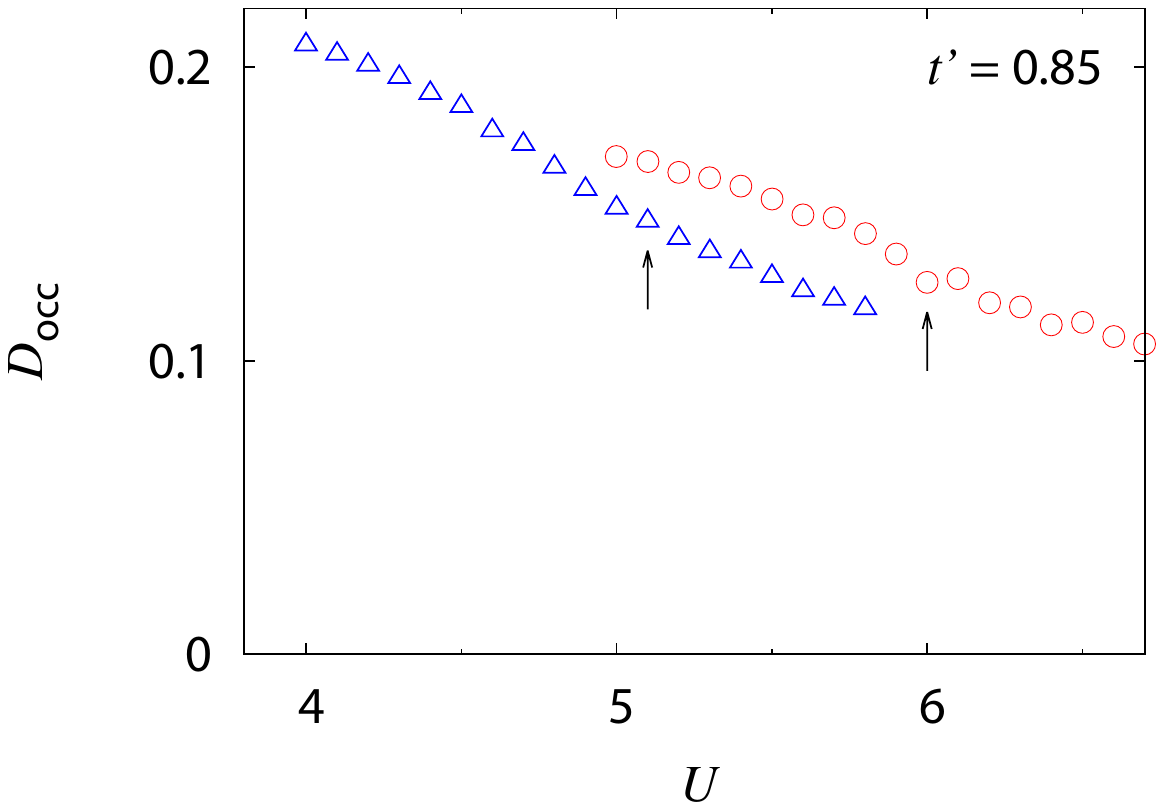}
\caption{
(Color online)
The double occupancies 
$D_{\rm occ}$ at $t' = 0.85$ computed on 12D cluster (circles) and the triangular 6-site cluster (triangle) as functions of $U$ 
assuming that no magnetic order is allowed. The arrows indicate the Mott transition points.
\label{fig:gap-docc}\\[-2.5em]
}
\end{figure}

Next we compare our results with the PIRG\cite{kawakami} and VMC\cite{watanabe,becca} studies. 
In the PIRG study\cite{kawakami} on the isotropic triangular lattice ($t'/t =1$), 
$U_{\rm MI} \sim 7.4 $ and $U_{\rm c} \sim 9.2$, 
which are slightly larger than our values. 
In the older VMC analysis\cite{watanabe} the region $0 \leq t' \leq1.2$ 
and $0 \leq U \leq 25$ was studied and non-magnetic insulator phase was not found. 
The later VMC study\cite{becca} reported that non-magnetic insulator is not realized at the isotropic point $t' = 1$, 
but it emerges around $t' \simeq 0.85$ for $U \simeq 12$ above the magnetic phase and 
covers the region $0.74 \lesssim t' \lesssim 0.98$ at $U \simeq 30$. 
In these VMC analyses\cite{watanabe,becca} $U_{\rm MI}$ is larger than ours (e.g., $ 8 \lesssim U_{\rm MI} $ at $t' = 1$ in them), 
and our non-magnetic insulator phase below the magnetic states in Fig.~\ref{fig:phase-diagram} is metal 
in their phase diagrams. 
Our results in Fig.~\ref{fig:order-u30-15} are qualitatively consistent with 
the recent VMC study\cite{becca} in the sense that non-magnetic insulator 
emerges above the magnetic states around $t' \simeq 0.9$, 
but this non-magnetic insulator phase in the VMC\cite{becca} is significantly wider 
than the modified spin wave theory\cite{hauke} and ours. 
\begin{figure}
\includegraphics[width=0.47\textwidth,bb = 146 328 468 514]{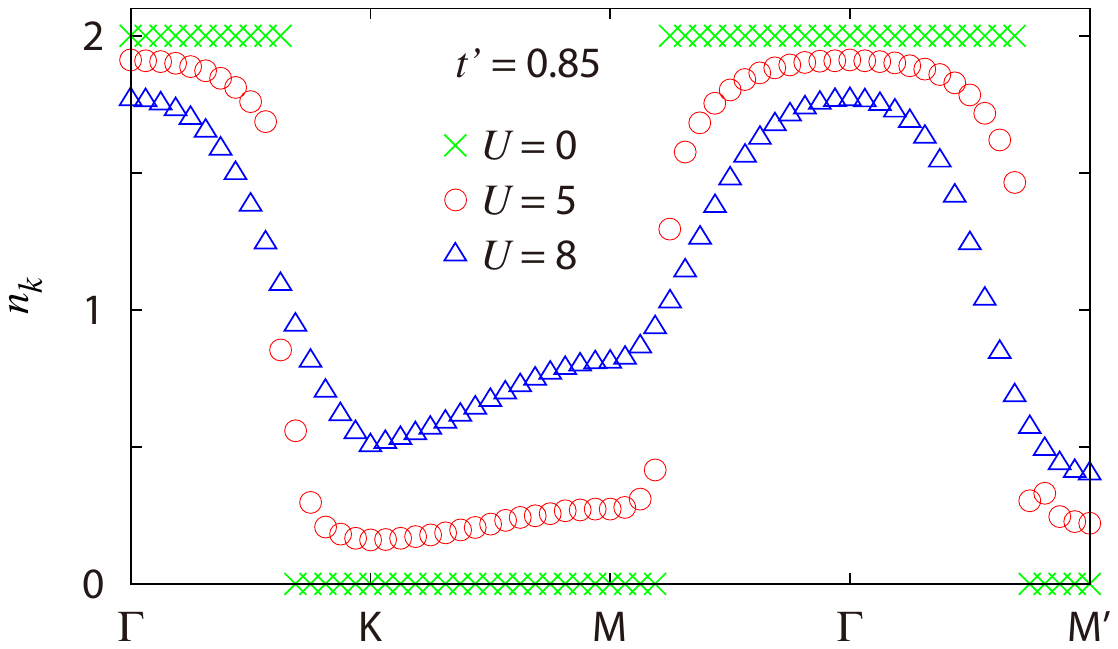}
\caption{
(Color online)
The momentum distribution function $n_k$ at $t' = 0.85$ along the dotted line in Fig.~\ref{fig:model}(b) 
for $U=0$ (crosses), $U=4$ (circles), and $U=8$ (triangles). 
\label{fig:nk}\\[-2.5em]
}
\end{figure}

\begin{figure*}
\includegraphics[width=0.94\textwidth,bb = 30 124 560 623]{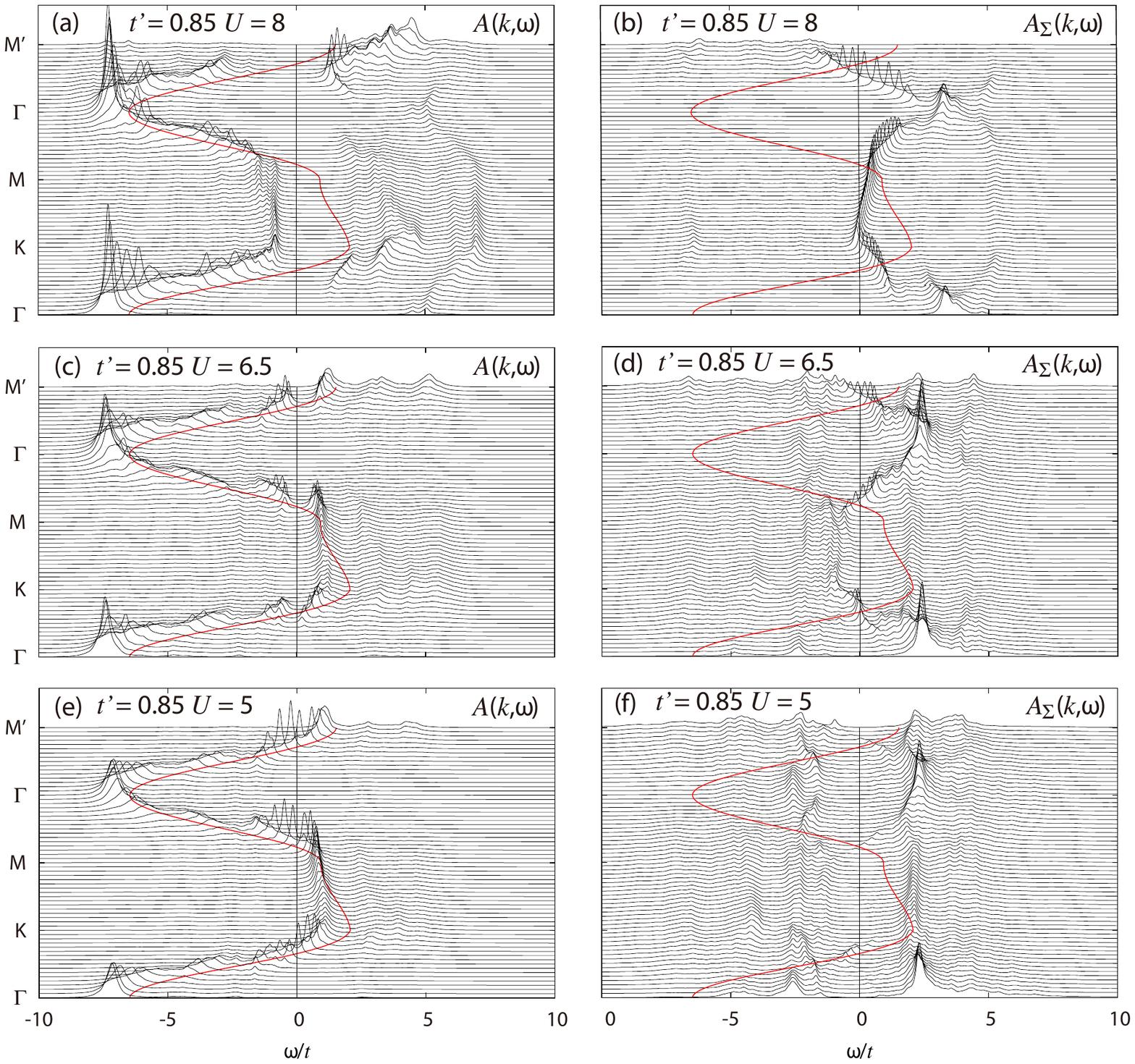}
\caption{
(Color online)
The spectral functions $A(k,\omega)$ and $A_{\Sigma}(k,\omega)$ calculated 
with $\eta = 0.1$ at $t'=0.85$ for 
(a) (b) $U=8$, (c) (d) $U=6.5$, and (e) (f) $U=5$, assuming that no magnetic order is allowed. 
The full lines are the non-interacting ($U=0$) band structure. 
\label{fig:sp-u5-7v2}
}
\end{figure*}

As for the disordered state of $\kappa$-(BEDT-TTF)$_2$Cu$_2$(CN)$_3$, 
an accurate estimation of $t'/t$ and $U/t$ seems to be very difficult. 
For example, $t'/t \simeq 1.1$ in Ref.~\onlinecite{salt-parameter1} while 
$t'/t \simeq 0.83 $ in Refs.~\onlinecite{salt-parameter2} and~\onlinecite{salt-parameter3}, and 
$5 \lesssim U/t \lesssim 8$ in Refs.~\onlinecite{salt-parameter1} and~\onlinecite{salt-parameter2} while 
$12 \lesssim U/t \lesssim 15$ in Ref.~\onlinecite{salt-parameter3}. 
If $U/t \simeq 15$, the spin liquid state of this material 
corresponds to the non-magnetic insulator above the magnetic states in Fig.~\ref{fig:phase-diagram} or \ref{fig:order-u30-15}. 
Otherwise, it corresponds to the non-magnetic insulator above the metal in Fig.~\ref{fig:phase-diagram}. 
These qualitatively different situations can be distinguished in experiments to see if the spin liquid 
state changes to the metal or magnetic state by applying the pressure and moving the system downwards in the phase diagram.

\section{Mott transition}

Here we study the Mott transition in detail. 
Fig.~\ref{fig:gap-docc} shows the double occupancies 
$D_{\rm occ}$ at $t' = 0.85$ computed on 12D cluster (circles) and the triangular 6-site cluster (triangle) as functions of $U$ 
assuming that no magnetic order is allowed (i.e. $h_{\rm M} = 0$). The arrows indicate the Mott transition points. 
The double occupancy is smooth at the transition point and 
the slope $ d D_{\rm occ}/dU $ around the transition point stays almost the same as the cluster 
size increases, thus tendency toward the first order transition was not observed.

In general, Mott transitions are predicted to be of the second order in 
VCA\cite{atsushi2011,atsushi2013} while they are predicted to be of the 
first order in the variational cluster approach with bath degrees of freedom. 
In the latter approaches, hybridization between the bath sites and cluster sites is treated 
as a variational parameter\cite{Potthoff2,Potthoff3} and the metal and insulator solutions in the coexisting region of the 
first order transition are different by the values of this hybridization parameter. 
VCA analyses do not have bath degrees of freedom and this technical difference may be the origin of the discrepancy. 

Next we analyze the spectral weight functions and related quantities. 
Fig.~\ref{fig:nk} shows the momentum distribution function $n_k$ at $t' = 0.85$ along the dotted line in Fig.~\ref{fig:model}(b) 
for $U=0$ (crosses), $U=4$ (circles), and $U=8$ (triangles), calculated imposing $h_{\rm M} = 0$. 
The clear steps in $n_k$ in the metal ($U=0,5$) disappear in the Mott insulator ($U=8$). 

We further study in detail the spectral density of the self-energy, and 
demonstrate that single dispersion evolves in the spectral representation of the self-energy around the Mott transition point 
and this dispersion gives rise to the Mott gap. 
Before showing the numerical results, we briefly discuss the relation between the Mott gap and dispersions in the self-energy 
using its spectral representation. 

The usual spectral density is defined by the Green's function $G(k\sigma,z)$ as 
\begin{eqnarray}
A(k\sigma,\omega) = - \frac{1}{\pi} {\rm Im} G(k\sigma, \omega  + i \eta), 
\end{eqnarray}
where $G(k\sigma,z)$ is expressed as 
\begin{eqnarray}
G(k\sigma, z ) = \frac{1}{z -  (\varepsilon_{k} -\mu)  - \Sigma(k\sigma,z) }
\label{eqn:g-sigma}
\end{eqnarray}
in terms of the free band $\varepsilon_{k}$, $\mu$, and the self-energy $\Sigma(k\sigma,z)$. 
The self-energy is expressed in the spectral representation\cite{l1} as
\begin{eqnarray}
\Sigma(k\sigma,z) = g_{k\sigma} + \int_{- \infty}^{\infty} 
\frac{\sigma_{k\sigma}(\xi)}{z-\xi} d \xi,
\,\,\,\,\,\sigma_{k\sigma}(\xi) \geq 0.
\label{eqn:sigma-spectral-rep}
\end{eqnarray}
and its spectral density is defined by 
\begin{eqnarray}
A_{\Sigma}(k\sigma,\omega) = - \frac{1}{\pi} {\rm Im} \Sigma(k\sigma, \omega  + i \eta).
\end{eqnarray}
For the Hamiltonian given by Eq. (\ref{eqn:hm}), 
\begin{eqnarray}
g_{k\sigma} = U\langle n_{-\sigma } \rangle,\,\,
\int_{- \infty}^{\infty} \sigma_{k\sigma}(\xi) d \xi = U \langle n_{-\sigma } \rangle (1 - \langle n_{-\sigma } \rangle),
\nonumber
\end{eqnarray}
where $\langle n_{\sigma } \rangle $ is the average number per site of particles with spin $\sigma$ 
in the ground state.\cite{eder1} Here we consider the paramagnetic state at half-filling 
$\langle n_{\pm \sigma } \rangle = 1/2$ and set $\mu = U/2$ assuming that $U$ is large. 
When the spectral weight $\sigma_{k\sigma}(\xi)$ is dominated by 
single pole of dispersion $\xi_k$, the Green's function is given by  
\begin{eqnarray}
G(k\sigma, z ) = \frac{1}{z - \varepsilon_{k} - \frac{ U^2/4}{z-\xi_k}  },
\label{eqn:g-sigma-approx}
\end{eqnarray}
where $g_{k\sigma}$ in $\Sigma(k\sigma,z)$ is canceled by $\mu$. 
The poles of this Green's function are 
$\omega_{\pm} = ( \varepsilon_{k} + \xi_k  \pm U)/2 $
for $ U \gg |\tilde{\varepsilon}_{k} |, |\xi_k | $, therefore 
the original band $\varepsilon_{k}$ splits into the upper and lower Hubbard bands 
(of almost equal weights) with a gap of width $U$. 

Fig.~\ref{fig:sp-u5-7v2} shows 
the spectral functions $A(k,\omega)$ and $A_{\Sigma}(k,\omega)$ calculated on 12D cluster 
with $\eta = 0.1$ at $t'=0.85$ for (a) (b) $U=8$, (c) (d) $U=6.5$, and (e) (f) $U=5$, 
assuming that no magnetic order is allowed. 
The non-interacting band is plotted with the full lines. 
At $U = 5$, which is slightly below the Mott transition point, 
small weights appear in $A_{\Sigma}(k,\omega)$ whenever the non-interacting band (full line) crosses the Fermi energy, 
as is seen in (f). 
At $U =6.5$, which is slightly above the Mott transition point, these weights grows and begins to form single dispersion in (d). 
Corresponding to this growth of the weights in $A_{\Sigma}(k,\omega)$ at the Fermi surface, 
a small gap opens at the Fermi surface in $A(k,\omega)$ in (c). 
In fact, as was calculated in Fig.~\ref{fig:order-tp} (b), a small insulating gap opens simultaneously 
all over the whole Fermi surface at $U=6.5$. 
As $U$ increases further, this dispersion dominates in $A_{\Sigma}(k,\omega)$ as is shown in (b), 
and yields the splitting of the non-interacting band into the upper and lower Hubbard band in (a). 

\section{Summary and conclusions}

In summary we have studied the magnetic properties and Mott transition in the Hubbard model on the anisotropic triangular 
lattice by VCA and the phase diagram is analyzed at zero temperature and half-filling. 
We found six phases, AF-metal, AF-insulator, spiral, AFC, paramagnetic metal, and non-magnetic insulator, 
which is the candidate of spin liquid. 
AF metal is realized for $ t'/t \lesssim 0.6$ and $U/t \lesssim 4$ and direct transitions from paramagnetic metal to 
AF insulator take place for $0.6 \lesssim t'/t \lesssim 0.8$. 
For $0.8 \lesssim t'/t \lesssim 1.2 $ paramagnetic metal changes to non-magnetic insulator at $U/t \simeq 6$, 
thus the (purely paramagnetic) Mott transition takes place there, and this non-magnetic insulator become 
magnetic state at $U/t \simeq 8$. 
Around $ t'/t \simeq 1.2$, magnetic state (AFC or spiral) is realized above the paramagnetic metal, 
and it changes to non-magnetic states as $U$ increases. 

The detailed comparisons of our results with those of the previous VCA study\cite{senechal} 
on the isotropic triangular lattice ($t'/t =1$) indicate that 
the main features of our phase diagram Fig.~\ref{fig:phase-diagram} remain almost the same in the thermodynamic limit. 

In our analysis, the three magnetic orderings AF, spiral (of 120$^\circ$ pitch angle), 
and AFC are considered to investigate non-magnetic states, and 
we can not exclude the possibility that magnetic orderings not approximated well by these orderings, e.g., 
magnetic orderings whose modulation period does not fit into the cluster size of VCA, 
are realized in the non-magnetic phase found in our study. 

As for the Mott transition, the structure of the self-energy in the spectral representation is studied in detail. 
As $U$ increases around the Mott transition point, single dispersion evolves in the spectral weights of the self-energy, 
which yields the splitting of the non-interacting band into the upper and lower Hubbard bands near the Fermi level. 

\vspace*{0.2cm}
{\it Note added.} After submitting the paper, a related VCA study of the Hubbard model 
on the anisotropic triangular lattice by M. Laubach {\it et. al}\cite{laubach} appeared on the preprint server. 
In that paper, the phase diagram is analyzed by VCA on the triangular 6-site cluster, which 
correspond to our Fig.~\ref{fig:phase-diagram-6}, and the energetically disfavored solutions were not shown in that study. 
They also computed the Mott transition point at $t'/t=1$ also on 12D cluster. 
Our numerical results agree with each other on the above points. 
In our paper, the results of the 6-site VCA (Fig.~\ref{fig:phase-diagram-6}) was included during the revision 
for the systematic study of the cluster size dependence of our 12D phase diagram.
\vspace*{0.2cm}


\section*{ACKNOWLEDGMENT}

The author would like to thank R.~Eder, K.~Harada, T.~Inakura, J.~Kokalj, H.~Kurasawa, H.~Nakada, T.~Ohama, Y.~Ohta, K.~Seki, 
and H.~Yamamoto for useful discussions. Parts of numerical calculations were done using the computer facilities of 
the IMIT at Chiba University and Yukawa Institute.

\end{document}